\documentstyle[12pt]{article}
\begin{document}
\begin{flushright}
UR-1453
\end{flushright}
\begin{center}
Comment on ``Quantum Phase of Induced Dipoles Moving in a Magnetic 
Field''\\
by\\
C. R. Hagen\\
Department of Physics \& Astronomy\\
University of Rochester\\
Rochester, NY  14627
\end{center}

\bigskip
\begin{abstract}

It has recently been suggested that an Aharonov-Bohm phase should 
be capable of detection using beams of neutral polarizable 
particles.  A more careful analysis of the proposed experiment 
suffices to show, however, that it cannot be performed regardless
of the strength of the external electric and magnetic fields.
\end{abstract}

It has recently been suggested [1] that an Aharonov-Bohm phase should 
be capable of detection using beams of neutral polarizable 
particles.  A more careful analysis of the proposed experiment 
suffices to show, however, that it cannot be performed regardless
of the strength of the external electric and magnetic fields.

To demonstrate this result one begins with the Lagrangian of ref. 1
$$L = {1\over 2} M {\bf V}^2 + {1\over 2}\alpha ({\bf E} + {\bf V} 
\times {\bf B})^2\eqno(1)$$
where $\alpha$ is the (intrinsically positive)
 polarizability of the particle and $M$ is its 
mass.  The electric field ${\bf E}$ is taken to have a magnitude $k/r$ in a 
radial direction in a plane perpendicular to the uniform magnetic 
field ${\bf B}$.  Upon applying the canonical formalism one readily 
obtains from Eq. (1) the relevant Hamiltonian in the form
$$H = {1\over 2} (M + \alpha B^2)^{-1}({\bf p} + \alpha B {\bf\bar 
E})^2 - {1\over 2} \alpha {\bf E}^2$$
where (as in ref. 1) motion has been restricted to the plane 
perpendicular to the magnetic field.  The two dimensional vector 
$\bar{\bf E}$ is the dual of ${\bf E}$ [i.e., $(\bar{\bf E})_i = 
\epsilon_{ij}E_j].$

One readily 
finds that the relevant Schrodinger equation for a particle of 
energy ${\cal E}$ is
$$\left[ {1\over 2}\left(M+\alpha B^2\right)^{-1}
 \left(-i\hbar \mbox{\boldmath $\nabla$}
+ \alpha B {\bar{\bf E}}\right)^2 - {1\over 2}\alpha{\bf E}^2\right]
\psi = {\cal E}\psi.$$
Standard separation of variables then yields for the radial wave 
function $f_m(r)$ the result

$$\left[{1\over r} {\partial\over \partial r} r 
{\partial\over\partial r} + {2(M+\alpha B^2){\cal E}\over \hbar^2} - 
{m^2 + 2m\alpha k B/\hbar - M \alpha k^2/\hbar^2\over r^2}\right] 
f_m(r)=0\eqno(2)$$
where $m = 0, \pm 1, \pm 2, \ldots$.

The system described by Eq. (2) is one which allows quantum 
mechanically well-defined solutions only when the numerator of the 
$1/r^2$ term is positive for all $m$.  This condition is readily
seen to be violated in the case of $m=0$ whenever there is a nonvanishing
electric field present.  Moreover, the case ${\bf E} = 0$ is clearly 
trivial in that it implies the absence of a quantum phase with the
only effect of the interaction being a mass renormalization (i.e.,
$M \to M + \alpha B^2)$.

In sum, the proposed experiment cannot be carried out because its 
assumptions are in basic conflict with quantum mechanics.
\bigskip

\noindent\underline{Acknowledgment}  

This work was supported in part 
by grant DE-FG02-91ER40685 from the U.S. Dept. of Energy.

\noindent\underline{References}

\noindent 1.  H. Wei, R.Han, and X. Wei, Phys. Rev. Lett. {\bf 75}, 
2071 (1995).

\end{document}